\documentclass[a4paper,showpacs,twocolumn,doublespace,prb,superscriptaddress]{revtex4}

\usepackage{graphicx}
\usepackage{dcolumn}
\usepackage{bm}
\usepackage{color}
\usepackage{ulem}
\expandafter\ifx\csname package@font\endcsname\relax\else
 \expandafter\expandafter
 \expandafter\usepackage
 \expandafter\expandafter
 \expandafter{\csname package@font\endcsname}%
\fi

\DeclareRobustCommand\substyle{\name@idx{document substyle}}%
\DeclareRobustCommand\classoption{\name@idx{document class option}}%
\DeclareRobustCommand\classname{\name@idx{document class}}%
\def\name@idx#1#2{%
 {\ttfamily#2}%
 \index{#2\space#1=\string\ttt{#2}\space#1}\index{#1>#2=\string\ttt{#2}}%
}%

\begin{document}
\title{Hydrogen site occupancy and strength of forces in nano-sized metal hydrides}
\author{Gunnar K. P\'alsson}
\author{Moritz W\"alde}
\affiliation{Division of Materials Physics, Department of Physics and Astronomy, Uppsala University, Box 516, S-75120 Uppsala, Sweden}

\author{Martin Amft}
\affiliation{Division of Materials Theory, Department of Physics and Astronomy, Uppsala University, Box, 516, S-75120 Uppsala, Sweden}

\author{Yuanyuan Wu}
\author{Martina Ahlberg}
\author{Max Wolff}
\affiliation{Division of Materials Physics, Department of Physics and Astronomy, Uppsala University, Box 516, S-75120 Uppsala, Sweden}
\author{Astrid Pundt}
\affiliation{Institut f\"ur Materialphysik, Universit\"at G\"ottingen, Hospitalstra\ss e 3-7, D-37073 G\"ottingen, Germany}
\author{Bj\"orgvin Hj\"orvarsson}
\affiliation{Division of Materials Physics, Department of Physics and Astronomy, Uppsala University, Box 516, S-75120 Uppsala, Sweden}

\date{\today}

\begin{abstract}
The dipole force components in nano-sized metal hydrides are quantitatively determined with curvature and x-ray diffraction measurements. Ab-initio density functional theory is used to calculate the dipole components and the symmetry of the strain field. The hydrogen occupancy in a 100 nm thick V film is shown to be tetrahedral with a slight asymmetry at low concentration and a transition to octahedral occupancy is shown to take place at around 0.07 [H/V] at 360 K. When the thickness of the V layer is reduced to 3 nm and biaxially strained, in a Fe$_{0.5}$V$_{0.5}$/V superlattice, the hydrogen unequivocally occupies octahedral $z$-like sites, even at and below concentrations of 0.02 [H/V].

\end{abstract}

\maketitle

\section{Introduction}
Hydrogen and its isotopes can be absorbed in large quantities into the interstitial sites of many transition metals.\cite{G:1978ue,Anonymous:1978ta}
Hydrogen constitutes the simplest impurity problem and exhibits many properties that can be explained by a lattice gas model.~\cite{MANCHESTER:1976vm,Alefeld:1969ut}
The interaction between the absorbed hydrogen atoms is of special interest, since it is mediated by the host metal through elastic deformations. \cite{Buck:1972wi,Alefeld:1972wf} For example, the long ranged attractive interaction is responsible for the condensation into liquid- and solid like phases. Understanding and quantifying the forces exerted by hydrogen is thus crucial for the understanding of phase transitions in metal hydrides. Due to the long range of the inter-hydrogen attraction in e.g. the transition metals, the phase formations additionally depend crucially on the elastic boundary conditions. Thus when studying hydrogen in nano-sized systems, large changes are expected when the aspect ratio and shape are altered.

Measurements of the hydrogen induced local elastic stress have been carried out in bulk metal systems~\cite{Fukai:1983if,G:1978ue} by measuring the components of the dipole force tensor, which is defined by
\begin{equation}
P_{ij}=\sum_{m}f_i^mR_j^m=\frac{\Omega}{c}\bar\sigma_{ij},
\label{stress_definition}
\end{equation}
where $R_{i}$ is the $i$-th component of the position of the $m$-th metal atom and $f_j^m$ is the $j$-th component of the force exerted by hydrogen on metal atom $m$. 
When hydrogen atoms (concentration $c$ in atomic ratio [H/M]) are distributed randomly in the host structure, the average stress $\bar\sigma_{ij}$ can be related to the components of the dipole tensor and the atomic volume of the metal ($\Omega$). 
The dipole force tensor is the connecting quantity between the atomistic and the elastic-continuum description of lattice distortions originating from point defects in crystals.~\cite{ELSASSER:1994wb} The dipole force tensor is analogous to an electric or magnetic dipole and has been used for instance in the description of ferroelasticity~\cite{Alefeld:1969vu}.

In bcc metals two distinct high-symmetry sites can be occupied by hydrogen, tetrahedral ($T$) and octahedral ($O$). These sites exhibit tetragonal symmetry and the dipole tensor can be written for the case of $O_z$ or $T_z$ sites as
\begin{equation}
P_{ij}=\left( \begin{array}{ccc}
B & 0 & 0  \\
0 & B & 0  \\
0 & 0 & A  \\
\end{array} \right).\label{dipole_stress}
\end{equation}

The dipole force tensor for other hydrogen absorption sites are straightforward permutations of Eq.~\ref{dipole_stress}. 
In general, the components of the dipole tensor are different for hydrogen occupying the two sites and thus serve as a signature for the type of site occupied.

The hydrogen-induced stress field in nano systems has previously been investigated by Kirchheim\cite{Laudahn:1999ud,Dornheim:2003cu,Laudahn:1999bu} and co-workers, where they showed, for instance, that the elastic constants of a Nb film could be determined from measurements of stress and hydrogen concentration. Stress measurements have also been used to follow the formation of different hydride phases in polycrystalline films of YH$_x$.~\cite{Dornheim:2003cu} Hydrogen absorbed into low-dimensional structures such as epitaxial thin films and superlattices can experience
different values of the dipole components due to strain effects. For example Fe/V superlattices are tetragonally strained with a ratio that depends on the ratio of the constituent materials.\cite{Andersson:1997dm,Bjorck:2007gd,Palsson:2008db} The hydrogen absorption properties of thin films and superlattices, investigated in the past, have shown distinctly different properties than that of the bulk.\cite{Andersson:1999bt,Olsson:2005ii,Olsson:2005jk,Olsson:2001bz,Leiner:2002fm,Hjorvarsson:1997jf} However, it has never been convincingly demonstrated into what site hydrogen goes in these types of structures nor to what extent the finiteness of the extension dictates the evolution of the stress field.

We measure directly the dipole components in a vanadium film and a Fe$_{0.5}$/V$_{0.5}$ (001) superlattice, as a function of the hydrogen concentration and  determine the site occupancy in these types of structures. We will also determine the components of the dipole force tensor from first principles by calculating the Hellmann-Feynman forces using density functional theory. This allows us to verify the occupancy of the two distinct sites as a function of H concentration, which were experimentally determined from measurements of the dipole components. 

\section{Experimental details}

\subsection{Sample preparation and characterization}

The Fe$_{0.5}$V$_{0.5}$$/$V  (6/21) ML superlattice with 100 repetitions was grown on a 20$\times$20$\times$1 mm$^{3}$ polished single-crystal MgO $(001)$ substrate using UHV-based magnetron co-sputtering. The base pressure in the system was 2$\times$10${^{-7}}$ Pa prior to deposition with the dominating residual gas being hydrogen. The sample was deposited from targets of iron and vanadium onto the substrate held at 573 K. Argon, with purity better than 99.9999 $\%$ at a pressure of 0.27 Pa was used as the  sputtering gas. The deposition rates of Fe and V were 0.0134 nm s$^{-1}$ and 0.0143 nm s$^{-1}$, respectively. The first vanadium layer was directly deposited onto the MgO substrate and the superlattice was capped with a vanadium layer followed by a 10 nm palladium layer. The palladium was sputtered from a target with 99.99 \% purity after cooling the sample to room temperature. The palladium layer acts as a catalyst for the hydrogen dissociation and protects the underlying structure from oxidation. The 100 nm vanadium film was grown epitaxially on MgO (001) 10$\times$10$\times$0.5 mm substrate using the same recipe as above except a 5 nm palladium cap was grown on top of the film.

The 100 nm V film was measured with x-ray diffraction and has a coherence length of 50 nm in the $z$ direction and a mosaicity of 0.2 degrees. Thus the film can be viewed as a single-crystal since the coherence length is of the order of the film thickness. The Fe$_{0.5}$V$_{0.5}$/V superlattice structure has been characterized previously using x-ray and neutron reflectivity as well as x-ray diffraction and transmission electron microscopy.~\cite{Paacutelsson:2010tf} It was determined to be a single-crystal that showed no signs of changes in the unloaded XRD or reflectivity pattern during repeated cycles of hydrogenation. This is interpreted as an absence of subsequent structural degradation.
\subsection{Curvature measurements}

\begin{figure}[!ht]
\includegraphics[width=8.6 cm]{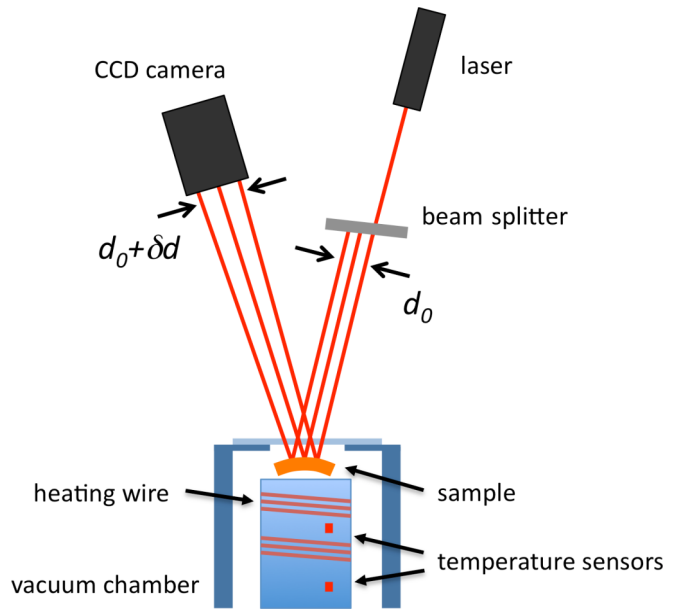}
\caption{A schematic illustration of the curvature measurement. A laser beam is split into a rectangular array of beams that reflect from a curved sample and is detected by a CCD camera. The change in spot spacing can be related to a change in curvature.}
\label{optical_setup}
\end{figure}

\begin{figure}[!ht]
\includegraphics[width=6.6 cm]{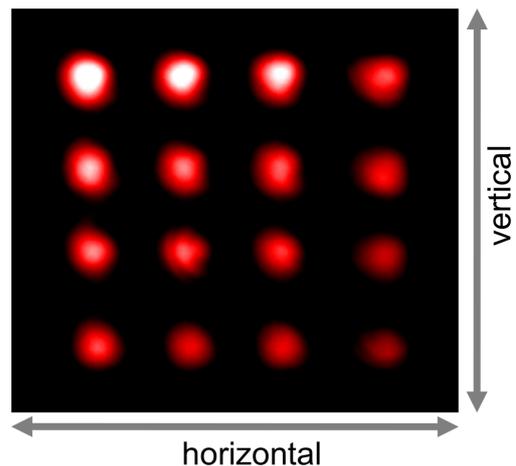}
\caption{A typical array of spots reflected from the sample surface and detected by the camera. The spot spacing yields the curvature according to Eq.~\ref{curvature_equation}.}
\label{spots}
\end{figure}

A multi-beam optical stress sensor (MOSS) is used to measure curvature in real-time, which is schematically depicted in Fig.~\ref{optical_setup}. A laser beam is split into multiple beams using two etalons, which reflect from the sample surface into a CCD camera. The whole laser system is enclosed and kept at a constant temperature to minimize the influence of thermal drift on the measurements. Commercial software (kSpace associates) is used to analyze the reflected pattern and calculate the curvature. The curvature instrument includes an ultra-high vacuum compatible chamber, which can be heated to 400 K. Capacitive membrane gauges are used to measure the pressure in the range 0.01-10 Pa. Figure~\ref{spots} shows a typical image recorded by the CCD camera from the sample. The change in spot spacing $d$ is directly related to the curvature $\kappa$ by
\begin{equation}
\kappa-\kappa_0=\frac{\delta d}{d_0}\frac{\cos\alpha}{2L},
\label{curvature_equation}
\end{equation}
where $L$ is the distance from the laser to the sample and $\alpha$ is the angle of incidence (see Fig.~\ref{optical_setup}).
The curvature of the film is measured directly as a function of temperature and external H$_2$ pressure. The curvature of the substrate is related to the stress in the film by Stoney's formula adopted for substrates with a (001) orientation:
\begin{equation}
\sigma_{ij}=\frac{1}{s_{11}+s_{12}}\frac{t_{\mathrm{sub}}^2}{t_{\mathrm{film}}}\kappa,
\label{stoney}
\end{equation} 
where $s_{ij}$ are the compliance constants of the MgO substrate and $t_{\mathrm{sub}}$ and $t_{\mathrm{film}}$ are the thickness of the substrate and the film, respectively. A didactic derivation of this version of Stoney's formula can be found in Ref.~[\onlinecite{Janssen:2009bj}]. The stress calculated from Stoney's formula is positive when the substrate curves as shown in the right panel of Fig.~\ref{stress} (concave).~\cite{Janssen:2009bj} The resolution of the curvature measurement is approximately $2\times 10^{-4}$ m$^{-1}$ and a typical uncertainty in stress is of the order of 0.04 GPa.

\subsection{In situ X-ray diffraction}
A specially designed UHV scattering chamber was used for determining both the hydrogen induced expansion and structural changes in the samples. This scattering chamber allows in-situ exposure to hydrogen in a wide temperature and pressure range, and was mounted on a Bruker Discover D8 X-ray diffractometer equipped with a parallel X-ray beam (CuK$\alpha_1$ $\lambda$=0.15406 nm). The sample was exposed to highly purified H$_2$-gas at different pressures at constant temperature (0.01-10 Pa). Thermodynamic equilibrium was ensured by continuously monitoring the (002) Bragg peak for changes after a sudden pressure increase.

\subsection{Dipole components of a clamped film}\label{sec:dipole}
The strain propagating through the metal due to the dipole force tensor is given by elasticity theory as
\begin{equation}
\left(\begin{array}{c}
 \epsilon_{1} \\ 
 \epsilon_{2} \\
 \epsilon_{3} \\
 \epsilon_{4} \\ 
 \epsilon_{5} \\
 \epsilon_{6} 
 \end{array} \right)
 =
\left( \begin{array}{cccccc}
s_{11} & s_{12} & s_{12} & 0 & 0 & 0 \\
s_{12} & s_{11} & s_{12} & 0 & 0 & 0 \\
s_{12} & s_{12} & s_{11} & 0 & 0 & 0 \\
0 & 0 & 0  & s_{44} & 0 & 0 \\
0 & 0 & 0 & 0 & s_{44} & 0 \\
0 & 0 & 0 & 0 & 0 & s_{44} 
\end{array} \right)
\left(\begin{array}{c}
 \sigma_{1} \\ 
 \sigma_{2} \\
 \sigma_{3} \\
 \sigma_{4} \\ 
 \sigma_{5} \\
 \sigma_{6} 
 \end{array} \right)\label{elasticmatrix}
\end{equation}
where $\epsilon_{1}$ is shorthand notation for $\epsilon_{11}$ (Voigt notation), etc., $s_{ij}$ are the compliance constants of the crystal and have units of inverse of pressure. The compliance constants for vanadium were converted from [Ref. \onlinecite{Magerl:1976vx}] and are $s_{11}=6.821\times10^{-12}$ Pa$^{-1}$ and $s_{12}=-2.336\times10^{-12}$ Pa$^{-1}$. Any change in the compliance constants due to hydrogen absorption and temperature can be completely neglected in the concentration and temperature range investigated.~\cite{Magerl:1976vx} However, since the superlattice is tetragonally strained and contains V, which is affected by the proximity to Fe, the elastic constants will be slightly different compared to the pure V constants. This difference is neglected in the present study.

The strain in the different directions for a free sample can now be evaluated by inserting Eq.~(\ref{stress_definition}) and Eq.~(\ref{dipole_stress}) into Eq.~(\ref{elasticmatrix}), which gives:
\begin{equation}
\begin{array}{ll}
\epsilon_1 = & \frac{c}{\Omega}[s_{12}A+(s_{11}+s_{12})B]\\
\epsilon_2  =& \frac{c}{\Omega}[s_{12}A+(s_{11}+s_{12})B]\\
\epsilon_3 = & \frac{c}{\Omega}[s_{11}A+2s_{12}B].

\end{array}\label{strain}
\end{equation}
The volume change is then evaluated as 
\begin{equation}
\frac{\Delta V}{V}=\epsilon_1+\epsilon_{2}+\epsilon_{3}=\epsilon_{1}+2\epsilon_{2}=\frac{c}{\Omega}(s_{11}+2s_{12})(A+2B).
\label{freevolumechange}
\end{equation}
To understand what happens to the volume change when a film is clamped on a substrate, the following thought experiment can be invoked. Assume that the film fits perfectly on the substrate and is free to expand according to Eq.~(\ref{strain}). Next, we impose an imaginary external stress in the plane of the film, which completely negates the in-plane expansion. We do not know what stress is required but the strain has to be
\begin{equation}
\epsilon_{1}^{\mathrm{ext}}=\epsilon_{2}^{\mathrm{ext}}=-\epsilon_{1}=-\epsilon_{2}
\end{equation}
and the stresses
\begin{equation}
\sigma_{1}^{\mathrm{ext}}=\sigma_{2}^{\mathrm{ext}}\qquad \sigma_{3}^{\mathrm{ext}}=0.
\end{equation}

By using Eq.~(\ref{elasticmatrix}), we can immediately write down some general relations for this kind of application of stress:
\begin{equation}
\epsilon_3^{\mathrm{ext}}=2s_{12}\sigma_{1}^{\mathrm{ext}} \qquad \epsilon_{1}^{\mathrm{ext}}=(s_{11}+s_{12})\sigma_{1}^{\mathrm{ext}}.\label{relations}
\end{equation}
Combining the relations in Eq.~(\ref{relations}) yields 
\begin{equation}
\epsilon_3^{\mathrm{ext}}=2\frac{s_{12}}{s_{11}+s_{12}}\epsilon_1^{\mathrm{ext}}.
\end{equation}
The total volume change due to hydrogen in a clamped film is thus
\begin{equation}
\epsilon_3^{\mathrm{tot}}=\epsilon_3+\epsilon_3^{\mathrm{ext}}=\frac{c}{\Omega}\frac{(s_{11}-s_{12})(s_{11}+2s_{12})}{s_{11}+s_{12}}A.
\label{totclamp}
\end{equation}
The stress $\sigma_1^{\mathrm{ext}}$ can now be evaluated and is
\begin{equation}
\sigma_1^{\mathrm{ext}}=-\frac{c}{\Omega}\Big(B+ \frac{s_{12}}{s_{11}+s_{12}}A\Big).
\label{realstress}
\end{equation}

\subsection{Concentration conversion}
Sievert's law was used to convert the H$_2$ pressure $p$ and temperature to concentration
\begin{equation}
c=K^{-1}(T)\sqrt{\frac{p}{p_0}},
\end{equation}
where $K$ is the Sievert constant. For the 100 nm V film the constant was obtained from the literature on 50 and 10 nm films, which are consistent with bulk solubility.~\cite{Bloch:2010cf,GRIFFITHR:1972tp}

The Sievert constant for the Fe$_{0.5}$V$_{0.5}$/V superlattice was obtained by measuring the relative change in the resistivity as a function of the applied hydrogen pressure at different temperatures. The relation between resistivity changes and concentration was obtained from previous neutron reflectivity measurements.~\cite{Paacutelsson:2010tf} The value of the Sievert constant was determined to be $K_{\mathrm{FeV/V}}^{-1}=0.078$ at 360 K.
\begin{figure}[!ht]
\begin{center}
\includegraphics[width=8.6 cm]{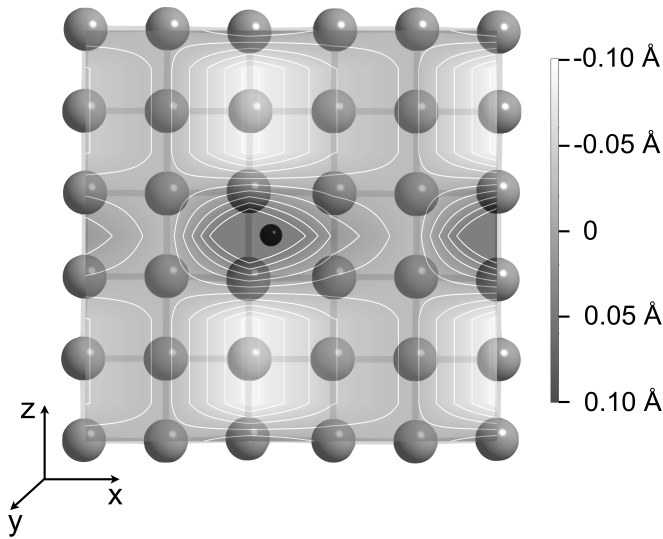}
\caption{ Illustration of the lattice distortion induced by a hydrogen atom (small dark ball) at the interstitial tetrahedral site in bcc bulk vanadium (larger grey balls). The overlayer shows the magnitude of the displacement field as changes in the inter-vanadium bond length (white: contraction; black: expansion). The contour lines of constant distortion are shown as white lines. The far field expansion is isotropic with H at the tetrahedral site. }
\label{displ_tetra}
\end{center}
\end{figure}

\subsection{Theoretical methods}
The scalar-relativistic  \textit{ab-initio} density functional theory (DFT)  calculations were performed using the projector augmented wave (PAW) \cite{Blochl:1994p10844,Kresse:1999p10843} method as implemented in \textsc{vasp}. \cite{Kresse:1996p6093,Kresse:1996p6092}
The exchange-correlation interaction was treated in the generalized gradient approximation (GGA) in the parameterization of Perdew, Burke, and Ernzerhof (PBE).\cite{PERDEW:1996p6520}
A cut-off energy of 500 eV was used and a Gaussian smearing with a width of $\sigma$ = 0.05 eV  for the occupation of the electronic levels.
For the bulk calculations the \textit{3p}$^{6}$, \textit{3d}$^{4}$, and \textit{4s}$^{1}$ states of vanadium were treated as valence states.
In the superlattice calculation the \textit{3d}$^{4}$ and \textit{4s}$^{1}$ states of vanadium as well as the \textit{3d}$^{7}$ and \textit{4s}$^{1}$ states of iron were treated as valence states.

The bulk bcc vanadium was modeled by a $3 \times 3 \times 3$ (54 atoms) or a $4 \times 4 \times 4$ (128 atoms) supercell, respectively, with a calculated lattice constant of 3.007 \AA.
The iron - vanadium superlattice was modeled by repeating a $2 \times 2$ supercell, consisting of 7 monolayers (ML) of iron and 7 ML of vanadium, in the z-direction. 
In the experiments the superlattice was grown epitaxially on a magnesium oxide (001) surface. 
This compresses the in-plane lattice constant of vanadium by 1.61\%.
In the calculations, we simulated this relative change by starting from the calculated lattice constant of MgO (4.21 \AA), which leads to an in-plane lattice constant of V of 2.93 \AA.
Note: the calculated bulk lattice constant of V with the smaller set of valence states is 2.98 \AA. Since the alloy does not take up appreciable hydrogen at the temperatures and pressure used in this study, it was deemed sufficient to use pure Fe instead. The full size of the superlattice was too large to be modeled within a reasonable time frame so Fe/V 7/7 ML was chosen as a reasonable compromise between ease of computation and still being physically relevant.

A Monkhorst-Pack $\Gamma$-centered $4\times4\times4$ k-point mesh (36 k-points in the irreducible wedge of the Brillouin-Zone) was used for the structural relaxations of the $3 \times 3 \times 3$ bcc V supercell, a $3\times3\times3$ k-point mesh for the $4 \times 4 \times 4$ bcc V supercell, and a $7 \times 7 \times 3$ k-point mesh for the Fe/V superlattice. Spin-polarization was taken into account in the bulk calculations.
The relaxation cycle was stopped when the Hellmann-Feynman forces had become smaller than $5 \cdot 10^{-3}$ eV/{\AA}. 

\section{Results and discussion}
\begin{figure}[!ht]
\begin{center}
\includegraphics[width=8.6cm]{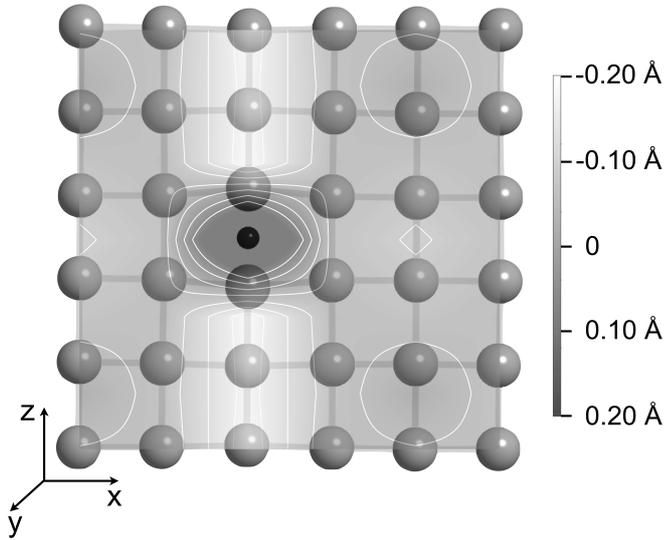}
\caption{Illustration of the lattice distortion induced by a hydrogen atom (small dark ball) at the interstitial octahedral site in bcc bulk vanadium (larger grey balls). The overlayer shows the magnitude of the displacement field as changes in the inter-vanadium bond length (white: contraction; black: expansion). Additionally, in white the contour lines of constant distortion are shown. Note that the expansion has a rotation symmetry with respect to an axis parallel to the z-axis that passes through H atom.}
\label{displ_octa}
\end{center}
\end{figure}

\subsection{Stress and strain}
Figure~\ref{displ_tetra} shows calculated local strain fields in bulk bcc vanadium with hydrogen occupying a tetrahedral $z$ site (T$_z$). The overlayer shows the magnitude of the displacement field as changes in the inter-vanadium bond length (white: contraction; black: expansion). Within the numerical accuracy, the displacement field is isotropic. Figure~\ref{displ_octa} shows the strain field with hydrogen occupying a octahedral $z$ site (O$_z$). Notice the strong asymmetry compared to the tetrahedral case. The forces exerted on the vanadium lattice before the lattice is allowed to relax can be used to approximate the dipole components according to Eq.~\ref{stress_definition}. The results are shown in Tab.~\ref{dipole_table} along with experimental values from the literature and experimental values from curvature and x-ray diffraction measurements. It is important to note that when extracting dipole components from experimental data, the elastic constants are assumed not to be affected. In other words, changes in the elastic compliance of the material due to hydrogen is incorporated into the values of the dipole components. However, the calculations yield, within the approximations of DFT, estimates of the physical, Hellmann-Feynman, forces.

\begin{figure}
\includegraphics[width=8.6 cm]{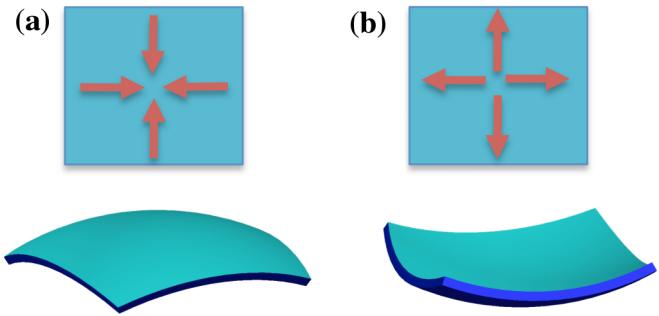}
\caption{Curvature changes due to film stress. The red arrows signify the stress experienced by the substrate due to the film. Panel (a) illustrates how compressive stress leads to convex curvature. Panel (b) shows that tensile in-plane stress results in concave curvature.}
\label{stress}
\end{figure}

A large discrepancy between calculated and measured dipole components from literature is found. To verify the computational approach, the calculations were repeated for one hydrogen atom in niobium and the results were in good agreement with previous calculations as well as experimental data. Furthermore, in the present calculation of H absorbed in bulk vanadium, a larger supercell, i.e. containing 128 atoms, and a larger set of valence states have been employed than in previous studies. Hence, our results are well-converged with respect to these parameters and the problem might arise from the vanadium pseudo-potentials as such. This conclusion is supported by a recent theoretical investigation~\cite{Koci:2008iu}, which also failed to correctly predict the elastic constants of vanadium at the GGA level. Even though the calculations fail to predict quantitatively the values of the dipole components, they qualitatively predict the correct symmetry of the strain field for the two different sites. The calculations show a decrease in the lattice spacing in the x-y directions when hydrogen occupies the $O_z$ site as shown in Fig.~\ref{displ_octa}, in agreement with experimental diffraction experiments.~\cite{Maeland:1964wf}

\begin{figure}
\begin{center}
\includegraphics[width=8.6 cm]{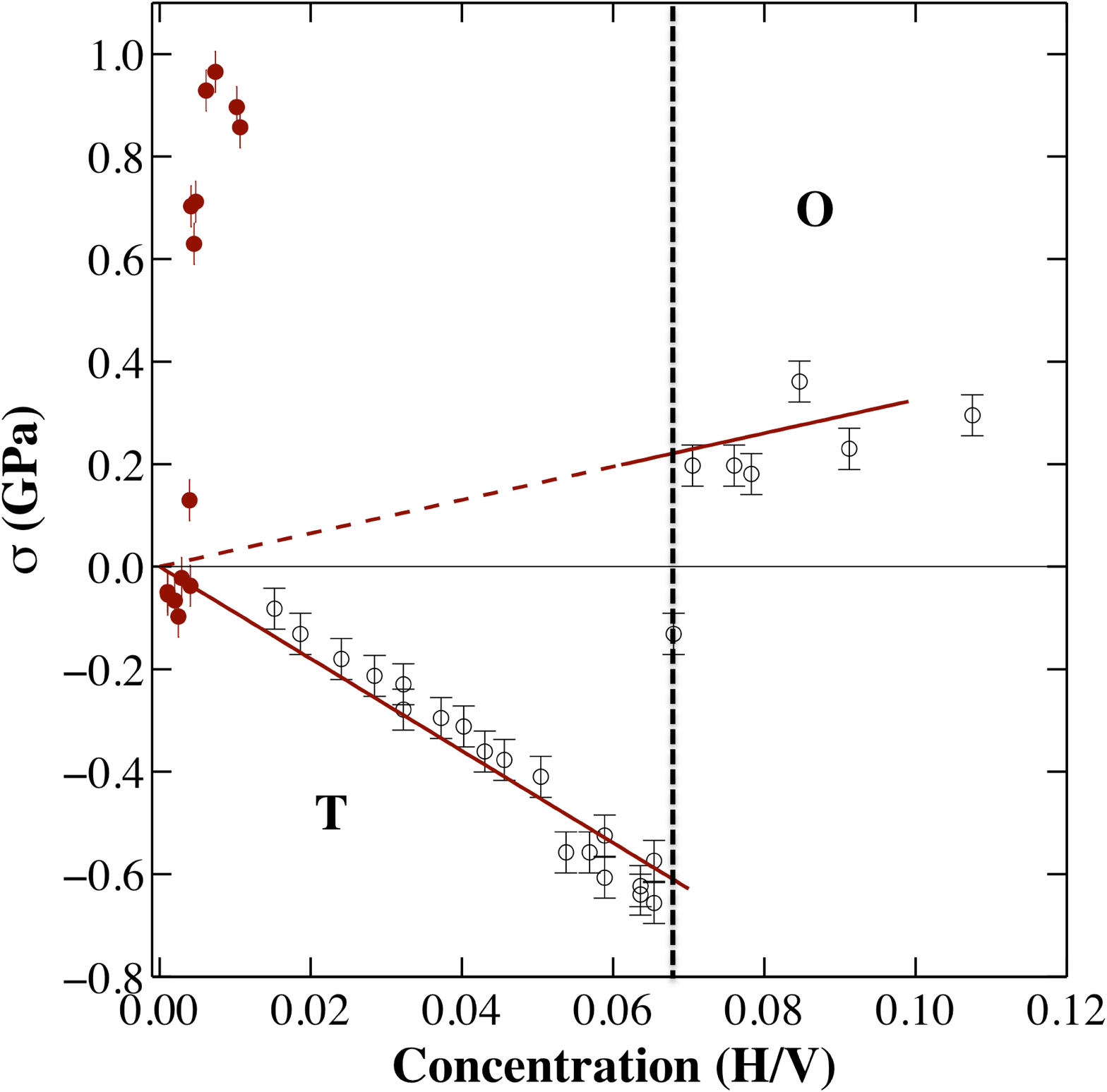}
\caption{In-plane stress as a function of concentration. The solid points correspond to the Fe$_{0.5}$V$_{0.5}$/V superlattice and the open points represent the 100 nm V film. The data is taken at 360 K.}
\label{conc_stress}
\end{center}
\end{figure}

We are now in a position to describe how hydrogen absorption into a film grown on a substrate dictates the curvature of the substrate. For a quantitative derivation see section~\ref{sec:dipole}. If hydrogen occupies the tetrahedral sites, the strain field propagates to the boundaries and contributes to an isotropic expansion. The substrate acts as a strong restoring force on the in-plane expansion, which leads to a compressive stress at the surface of the substrate. This in turn produces a bending moment and the substrate bends convex as shown in panel (a) of Fig.~\ref{stress}. Therefore, it is expected a priori that a flat sample, where hydrogen occupies tetrahedral sites, exhibits a convex curvature. When hydrogen occupies octahedral $z$ sites in the bulk, as found in the $\beta$-phase of VH$_x$, the material in the bulk contracts in the plane and expands out of plane. To fit the sample on a substrate the sample has to be stretched in the plane with a tensile stress. The restoring force of the substrate is thus tensile and the sample bends concave as shown in panel (b) of Fig.~\ref{stress}. Therefore it is expected that a flat sample with hydrogen occupying octahedral $z$ sites exhibits a concave curvature. 

\begin{figure}[!ht]
\begin{center}
\includegraphics[width =8.6cm]{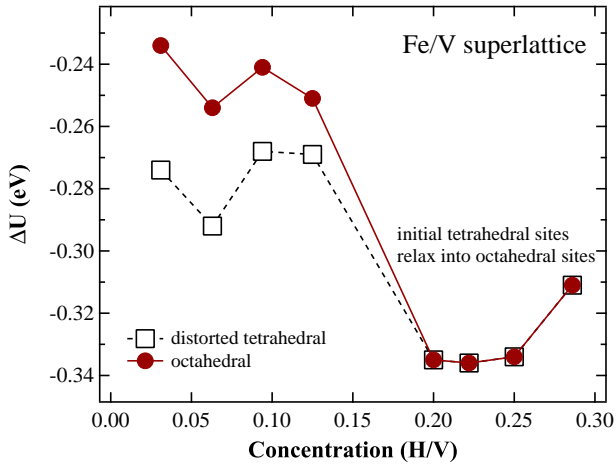}
\caption{Changes in the total energy of a Fe/V superlattice as a function of hydrogen concentration. The hydrogen atoms are absorbed at distorted tetrahedral (open squares) and octahedral (full circles) sites.}
\label{energy}
\end{center}
\end{figure}

Figure ~\ref{conc_stress} shows the in-plane stress of the 100 nm vanadium film (open circles) as a function of concentration. The stress is obtained by converting the measured curvature using Eq.~\ref{stoney}. A sharp change in the stress is observed at 0.065 in [H/V]. Below this transition the stress increases linearly with concentration as expected from the definition  Eq.~(\ref{stress_definition}). From the discussion above, we therefore assign the region below the transition to hydrogen occupying sites that have isotropic dipole components (tetrahedral sites). The region above the transition is where the dipole components are sufficiently different from each other to cause a contraction in the plane of the sample. Thus from a qualitative point of view, these results are consistent with tetrahedral occupancy at low concentrations and octahedral at concentrations above 0.065 [H/V], which resembles the bulk occupancies. 

Fig.~\ref{conc_stress} also shows the stress for a Fe$_{0.5}$V$_{0.5}$/V superlattice (filled circles). In stark contrast to the film results, we see an almost immediate increase (positive) in the stress with increasing concentration. There are therefore significant differences in the response from a film and a superlattice, which will be discussed more thoroughly below. 

\begin{table*}[ht!]
\renewcommand{\arraystretch}{1.8}
\begin{tabular}{ccccccccl}
\hline\hline
 & site & $A+2B$ [eV] & $\frac{3|A-B|}{A+2B}$ & $A$ [eV] & $B$ [eV] & $\frac{1}{c}\frac{\Delta V}{V}$ & $\frac{1}{c}\frac{\Delta V}{V}_{\mathrm{cl}}$ \\
 \hline
bulk V& T  & 7.7(4) & - & 2.6(1) & 2.6(1) & 0.189(7) & 0.129(7) & Magerl (1976) \\ 
bulk V& T  & 8.0(1) & - & 2.67(7) & 2.67(7)  & 0.197(4) & 0.134(3) & Schober (1988) \\ 
bulk V& T  & 7.4(2) & - & 2.47(5) & 2.47(5)  & 0.182(4) & 0.124(3) & Maeland (1964) \\ 
bulk V& T  & 7.2(6) & - & 2.4(2) & 2.4(2) & 0.18(1) & 0.12(1) & Schauman (1970) \\ 
100 nm V & - & 6.4(2) & 0.11 & 2.35(1) & 2.0(1) & - & 0.1189(7) & Present work\\
bulk V (calc.) & T & 11.26 & 0.09 & 3.54  & 3.86 &  & &Present work (V128H1 13 val) \\
\hline
bulk V& O$_z$ & 6.85 & 0.97 & 3.760 & 1.545 & 0.168 & 0.189 &  Fukai (1983) \\
100 nm V & O$_z$ & 7.2(3) & 0.9 & 3.8(1) & 1.7(2) & 0.17(1) & 0.19(1) & Present work \\
bulk V (calc.) & O$_z$ & 12.79 & 2.25  & 10.67  & 1.06 & 0.180 &  -&Present work (V128H1 13 val) \\
\hline

Fe$_{0.5}$V$_{0.5}$/V 6/21& - & - & - & 4.1(4)  &  - & - &  0.21(2) & P\'alsson (2010)\\

Fe/V 7/7 (calc) & T & 12.97 & 0.29 &  5.15 & 3.91 &  - & 0.105 &Present work \\ 
Fe/V 7/7 (calc) & O$_z$ & 13.31 & 1.95 & 10.21 & 1.55  & - & 0.140 &Present work \\

\hline
Nb (calc) & T & 10.45 &  0.07 & 3.31  & 3.57 &  & &Present work (Nb54H1) \\
Nb (calc) & T & 11.0 &  0.08 & 3.47  & 3.77 &  & & Sundell (2004)\\
Nb (calc) & O$_z$ & 13.2 & 2.11  & 10.6  & 1.3 &  & & Sundell (2004)\\
\hline\hline
\end{tabular}
\caption{Dipole force tensor components for vanadium hydride. The components are calculated from the trace and the anisotropy. The expansion is calculated using $A$ and $B$ and Eq.~(\ref{freevolumechange}). The clamped volume change is calculated from $A$ using Eq.~(\ref{totclamp}). Items labeled as (calc) are results from ab-initio DFT calculations.}
\label{dipole_table}
\end{table*}

Using the dipole components from literature, Eq.~\ref{realstress} predicts the stress on a MgO substrate to be $\sigma_1/c=4.9904$ GPa for octahedral $z$ sites and $\sigma_1/c=-14.0913$ GPa  for tetrahedral sites. We see that the two stresses have different signs, which is expected from the net contraction in the plane for hydrogen occupying octahedral sites and a net expansion for hydrogen occupying tetrahedral sites (see also Fig.~\ref{stress}). 

Table~\ref{dipole_table} shows literature values for the sum and the difference of the dipole components for tetrahedral sites $T$ and  $O_z$ sites. The volume change of a free sample can be calculated using Eq.~\ref{freevolumechange}. The volume change expected from a clamped film is also calculated from Eq.~\ref{totclamp}. Notice that the volume change is larger for the clamped film when hydrogen occupies $O_z$ sites than its free counterpart. In a previous publication~\cite{Paacutelsson:2010tf}, we used a simplified version of the above equation, which yielded a lower than expected value based on octahedral occupancy. Using the full elastic theory the results previously published can now be understood and are consistent with octahedral $z$ site occupancy. Note that the out of plane expansion does not depend on the in-plane components $B$. However the in-plane stress measured by the curvature change depends on both $A$ and $B$. By measuring the out-of-plane expansion the $A$ components can be uniquely determined, as described in the following chapter.

\subsection{Lattice expansion}

 \begin{figure}[!htpb]
\begin{center}
\includegraphics[width=8.6 cm]{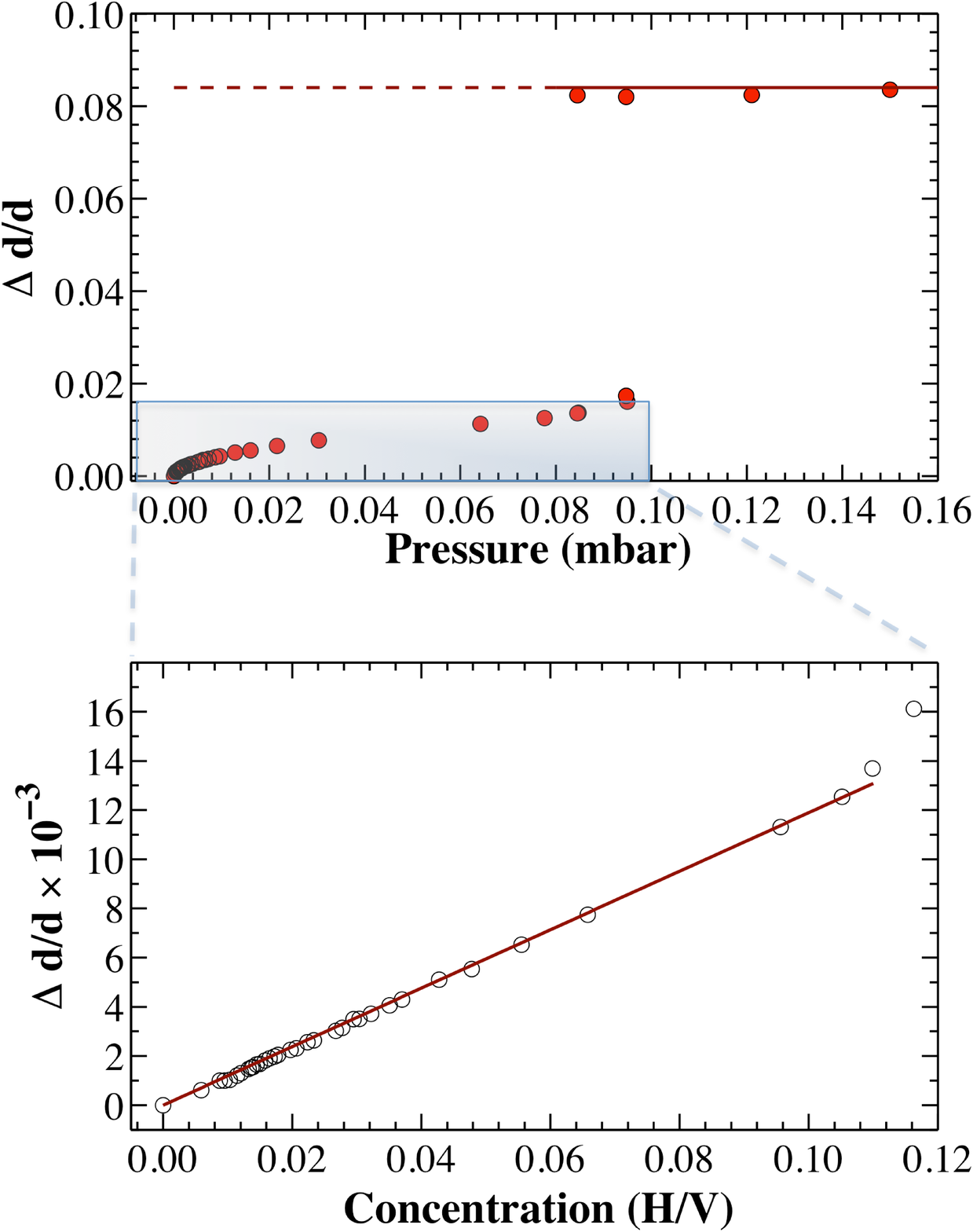}
\caption{Upper panel shows relative change in $d_{002}$ plane distance as a function of pressure. A co-existence region is identified in the region between 0.08-0.10 mbar. In the lower panel the region below the plateau is shown in more detail and with the pressure converted into concentration using Sievert's law.  The solid line is a linear fit to the data with the origin fixed at zero. The slope of the curve is 0.1189(7).}
\label{dd_vs_c}
\end{center}
\end{figure}

Figure ~\ref{dd_vs_c}. shows the relative change in the $d_{002}$ lattice distance as a function of pressure (upper panel) and concentration (lower panel) at 400 K for the 100 nm vanadium film. The lower panel shows the data converted from pressure to concentration in the region below the transition. The slope is 0.1189(7), which gives, using Eq.~\ref{totclamp}, a dipole component $A=2.35(1)$ eV. Using the value of $A$ obtained from the expansion and Eq.~\ref{realstress} we can determine $B$ by comparing to Fig.~\ref{conc_stress}. This is shown by the solid line in the concentration range 0.00-0.068 [H/V]. The component is slightly smaller than $A$ with a value $B=2.0(1)$ eV. The horizontal line in the upper panel of Fig.~\ref{dd_vs_c} gives using Eq.~\ref{totclamp} with $c=0.44$ a value $A=3.8(1)$ eV, which is consistent with octahedral occupancy. The concentration corresponds to the end of the isotherm plateau at 400 K. It is worth noting that large changes in pressure are necessary to change the concentration beyond the plateau, which is why a single concentration was used. For the region after the transition Eq.~\ref{realstress} was used assuming an asymmetry of the components with $A=3.8(1)$ eV as determined from diffraction and $B=1.7(2)$ eV.

It can thus be unambiguously concluded that the hydrogen occupies tetrahedral sites in 100 nm thick V films below the phase boundary and octahedral sites beyond the co-existence region. This film can be viewed as a consistency check for the methodology presented. The dipole components are in quantitative agreement with literature except for a slight tetragonality in the $B$ component for tetrahedral occupancy. This implies that Snoek anelastic relaxation~\cite{Buchholz:1973gq} could in principle be found for hydrogen in vanadium films, which has previously been excluded for the bulk V, Nb, and Ta hydrides.~\cite{Buchholz:1973gq}

Since an unusually large in-plane stress was observed for the Fe$_{0.5}$V$_{0.5}$/V superlattice, a fit to the data using Eq.~\ref{realstress} was not attempted. It is clear that although the sign from the stress measurements and the magnitude of the $A$ component determined from previous neutron studies~\cite{Paacutelsson:2010tf} point towards an octahedral $z$ site occupancy, we cannot explain the large stress observed. A calculation of the change in total energy of a 7/7 Fe/V superlattice with hydrogen (Fig.~\ref{energy}) shows that a distorted tetrahedral site is favored over the octahedral $z$ site at low concentrations. It is possible that the occupancy of this distorted site is related to the unusual stress. Higher concentration measurements of the stress are required to elucidate this possibility which however awaits further study. The volume expansion in clamped vanadium films is larger than the free counter part, when hydrogen occupies the octahedral sites. This is due to the large asymmetry of the dipole components, as described by continuum elasticity theory. The present methodology is not restricted to hydrogen in metals but works for all types of interstitial defects that can be quantified by a dipole tensor.

\section{Conclusions}
It is concluded that the force dipole components can be quantitatively measured by combining curvature and x-ray diffraction measurements in nano-systems. A 100 nm film of vanadium loaded with hydrogen exhibits a stress field that is similar to tetrahedral occupancy below 0.065 [H/V] at 360 K. Above 0.068 [H/V] a sharp transition to octahedral sites occurs, which is consistent with the known phase boundaries of VH$_\mathrm{x}$. Hydrogen in superlattices composed of Fe$_{0.5}$V$_{0.5}$/V are shown for the first time to occupy octahedral $z$-like sites at all measured concentrations. The symmetry of the strain fields due to hydrogen in different sites are qualitatively reproduced using {\it ab-initio} DFT calculations. The calculations also predict a change of site from tetrahedral to $O_z$ at low concentrations. However, the quantitative discrepancies between our experiment and calculations show the need to treat the inter-atomic interactions in metal-hydrides with methods beyond the GGA level. The methodology presented here opens up new routes for investigating quantitatively, the evolution of local stress fields due to point defects in nano-sized materials.

\section{Acknowledgements}
The Swedish research council ({\sc VR}) and Knut and Alice Wallenberg ({\sc KAW}) are acknowledged for financial support. M. Amft is grateful to the Swedish National Infrastructure for Computing (SNIC) for the granted computer time.
\bibliographystyle{apsrev.bst}
\bibliography{bibliography.bib}
\end{document}